\documentclass[12pt]{article}
\usepackage{graphicx}
\usepackage{mathrsfs}
\usepackage{amsmath,epsfig,epsf,psfrag}
\usepackage[nolists]{endfloat}
\usepackage{amscd}
\usepackage{amssymb}

\begin{document}
\title{The Limiting Distribution of Decoherent Quantum Random Walks}

\author{Kai Zhang\\
  Department of Mathematics, Temple University \footnote{Currently at the Department of Statistics, University of Pennsylvania.}\\
  E-mail: zhangk@wharton.upenn.edu
 }
\maketitle
\begin{abstract}
The behaviors of one-dimensional quantum random walks are strikingly
different from those of classical ones. However, when decoherence is
involved, the limiting distributions take on many classical features
over time. In this paper, we study the decoherence on both position
and ``coin'' spaces of the particle. We propose a new analytical
approach to investigate these phenomena and obtain the generating
functions which encode all the features of these walks.
Specifically, from these generating functions, we find exact
analytic expressions of several moments for the time and noise
dependence of position. Moreover, the limiting position
distributions of decoherent quantum random walks are shown to be
Gaussian in an analytical manner. These results explicitly describe
the relationship between the system and the level of decoherence.
\end{abstract}

\copyright 2007 United States Copyright Office

\newtheorem{notation}{Notation}[section]
\newtheorem{rem}{Remark}[section]
\newtheorem{lem}{Lemma}[section]
\newtheorem{cor}{Corollary}[section]
\newtheorem{tem}{Theorem}[section]
\newtheorem{prop}{Proposition}[section]
\newtheorem{example}{Example}[section]
\newtheorem{def1}{Definition}[section]
\newcommand{\bra}[1]{\langle #1 \vert}
\newcommand{\ket}[1]{\vert #1 \rangle}
\newcommand{\ip}[2]{\langle #1 \vert #2 \rangle}
\newcommand{\lo}[3]{\bra{#1} #2 \ket{#3}}

\section{Introduction}
Quantum random walks recently gained great interest from physicists,
computer scientists and mathematicians. The interest was sparked by
their important roles in developing highly efficient quantum
algorithms. For instance, Grover's search
algorithm \cite{GR96} has time cost $O(\sqrt{N})$, in contrast to 
the ordinary search algorithm which has a cost of $O(N)$. This
quantum search algorithm was proved to be closely related to the
behaviors of quantum random walks in \cite{ABD06} and \cite{SHE03}.
As another example, Shor's algorithm also improved the speed of
factorization dramatically \cite{SH97}. The high efficiency of
quantum algorithms is discussed in \cite{GR97}, \cite{GR98} and
\cite{RI06}. Experimental implementations of the algorithms are
discussed in \cite{DU02} and \cite{DU03}.

Besides their important applications, quantum random walks are very
attractive due to their dramatic non-classical behavior. After
quantum random walks were defined in \cite{AH93}, many articles
(\cite{AM01}, \cite{GR04}, \cite{KO052}, \cite{MA02}, \cite{MO02},
etc.) studying the distribution of quantum random walks were
published. It is known that the observed non-classical behavior is
due to quantum coherence \cite{RO04}. One of the most shocking
differences \cite{AM01} is that as time $t$ grows, the variances of
quantum random walks are $O(t^2)$ while the variances of classical
random walks are $O(t)$. Various limit theorems of quantum random
walks are established (\cite{GR04}, \cite{KO052}, \cite{GO05},
\cite{KO05}). An excellent reference can be found in \cite{KE03}.

One of the most important issues surrounding the use of quantum
random walks is that they are very sensitive to inevitable
decoherence, which could be caused by many reasons, such as
interactions with the environment and system imperfections
(\cite{BR03}, \cite{BR032}, \cite{SH03}). The effect of decoherence
is very important for the application of quantum algorithms, as
discussed in \cite{KEN03}, and \cite{RI07}. For the one-dimensional
case, in the model in \cite{BR03}, decoherence is introduced by
measurements on the particle's chirality. Long-term first and second
moments of the walk were obtained and numerical results showed that
the distributions look like classical normal distributions. Similar
results are found in other models (\cite{RO04}, \cite{KEN03},
\cite{JE04}, \cite{KEN02}, \cite{LO03}, \cite{RO05}, \cite{AB06}).
If we denote by $X_t$ the position of the particle obtained by
measurement of the wave functions of the decoherent quantum random
walk, all of above papers mentioned the fact that the variance of
the simulated $X_t$ grows linearly in time for large $t$ when the
quantum random walk is subject to decoherence.

These results stimulated us to prove that the one-dimensional
decoherent quantum random walk, $X_t \over \sqrt{t}$, converges to a
normal distribution. Our work focuses on the one-dimensional
discrete-time Hadamard walk with measurements taken on both position
and chirality at each time step. This kind of decoherence is studied
numerically in \cite{DU02}, \cite{KEN03}, \cite{JE04} and
\cite{KEN02} but we will study it fully analytically.

We shall see that when the particle is not measured, then the system
evolution is purely quantum and $X_t \over \sqrt{t}$ does not
converge. However, when the particle is measured subject to a small
probability, $X_t \over \sqrt{t}$ will converge to a normal
distribution. In the limit, when the particle is measured at each
step, then the system becomes purely classical and $X_t \over
\sqrt{t}$ is asymptotically standard normal.

In the next section, we introduce the mathematical setup of
decoherent quantum random walks. We then provide our methodology of
generating functions and the decoherence equation. We next list our
results and discuss the interesting phenomena that occur when $p$ is
small. Finally, we summarize and discuss our work. Mathematical
proofs are given in the appendix.

\section{Decoherent Quantum Random Walks}
\subsection{Pure Quantum Random Walk system}
We start with a brief description of the one-dimensional pure
quantum random walk system. In the classical random walks, the
particle moves to the right or left depending on the result of a
coin toss. However, in the quantum random walks, the particle has
its chirality $\{right, left\}$ as another degree of freedom. At
each step, a unitary transformation is applied to the chirality
state of the particle and the particle moves according to its new
chirality state.

We denote the \emph{position space} of the particle by $H_p$, the
complex Hilbert space spanned by the orthonormal basis $\{\ket{x}, x
\in \mathbb{Z}\}$, where $\mathbb{Z}$ is the set of integers. We
also denote the \emph{coin space} by $H_c$ as the complex Hilbert
space spanned by the orthonormal basis $\{\ket{l}, l = 1, 2\}$ where
$1$ stands for ``moving right'' and $2$ stands for ``moving left.''
The \emph{state space} $H$ of the particle is defined as
\begin{equation}
H = H_p \otimes H_c.
\end{equation}

A vector $\ket{\psi} \in H$ with $L^2$-norm 1 is called a
\emph{state} and tells us the distribution of the particle's
position and chirality upon measurements. The \emph{basis} of $H$
are denoted by $\{\ket{x,l} = \ket{x} \otimes \ket{l}: x \in
\mathbb{Z}\}, l=1,2\}$. Now we introduce the evolution operator
which drives the particle. The \emph{shift operator} $S: H
\rightarrow H$ is defined by
\begin{equation}
S \ket{x}\ket{l}= \left \{
\begin{array}{ll}
\ket{x+1}\ket{1} & l = 1\\
\ket{x-1}\ket{2} & l = 2.
\end{array}
\right .
\end{equation}
The \emph{coin operator} $C: H_c \rightarrow H_c$ can be any unitary
operator and is an analogue to the coin flip in the classical walk.
The \emph{evolution operator} $U: H \rightarrow H$ is defined by
\begin{equation}
U = S (I_p \otimes C),
\end{equation}
where $I_p$ is the identity in the position space.

Now let $\ket{\psi_0} \in H$ be the initial state and let
$\ket{\psi_t} = U^t \ket{\psi_0}$. The sequence
$\{\ket{\psi_t}\}_0^{\infty}$ is called a \emph{one-dimensional
quantum random walk}.

The most famous and best-studied example of quantum random walks is
the Hadamard walk, in which the coin operator is the $2 \times 2$
Hadamard matrix
\begin{equation}
H_2 = \frac{1}{\sqrt{2}} \left(
\begin{array}{cc}
1 & 1\\
1 & -1
\end{array}
\right).
\end{equation}
The quantum random walk associated with $H_2$ is called a
\emph{one-dimensional Hadamard walk}.

The \emph{probability} of a particle at state $\ket{\psi}$ to be
found at state $\ket{\eta}$ is defined by the norm squared of the
inner product of $\ket{\psi}$ and $\ket{\eta}$,
$|\ip{\psi}{\eta}|^2$.

In particular, the probability of the quantum random walk, starting
from the position $x=0$, with the coin in state $m$, to be found at
$x$ with coin state $n$ is
\begin{align}
W_{m,n}(x,t)=|\langle x, n \vert U^t \vert 0, m \rangle |^2.
\end{align}

\subsection{Decoherence}
We focus on decoherence caused by measurements that is on both
position and coin of the particle. A set of operators $\{A_i, i \in
\mathbb{A} \}$ is called a \emph{measurement} if
\begin{align}
\sum_{i \in \mathbb{A}} A_i^ * A_i = I,
\end{align}
where $\mathbb{A}$ is some index set and $A^*$ is the adjoint
operator of $A$, i.e., the complex conjugate of transposed matrix of
$A$.

In this work, we consider the measurements in a similar manner as in
\cite{BR03}. Let $p$ be a real number in $[0,1]$, to denote the
probability of the random walk being measured at each step. We let
$A_c : H \rightarrow H$ be s.t. $A_c = \sqrt{1-p} I$. Hence, $A_c$
is the coherence projection. We also let $A_{x,n} : H \rightarrow H$
be s.t. $A_{x,n} =\sqrt{p}\ket{x,n}\bra{x,n}$ to be the decoherence
projection to the subspace spanned by $\ket{x,n}$. Under this setup,
the index set $\mathbb{A}$ is $\mathbb{A}= \{c\} \bigcup \{(x,n): x
\in \mathbb{Z}, n=1, 2\}$.

Let $\ket{\psi}$ be a state in $H$. Then the position of the
particle, $X_t^{\psi}$ over $\mathbb{Z}$ obtained by taking
measurement of the wave function of the decoherent quantum random
walk starting from $\ket{\psi}$, at time $t$ has the following
distribution:

\begin{equation}\label{defdq}
\begin{split}
 & \!P(X_t^{\psi}=x)   \\
=& \sum_n  \sum_{j_n\in \mathbb{A}} \!\! \ldots \!\!\! \sum_{j_1\in
\mathbb{A}} \!\! | \lo{x,n} {(A_{j_t}U)\ldots(A_{j_1}U)}{\psi}|^2.
\end{split}
\end{equation}

In other words, the walk starts at $\ket{\psi}$, then we apply the
evolution operator $U$ and try to measure it. The process repeats
until the $t$-th step is finished. We then consider the position
distribution of the particle. We call each $(j_1, j_2, \ldots, j_t,
(x,n))$ a path. We also call $\lo{x,n}{
(A_{j_t}U)(A_{j_{t-1}}U)\ldots(A_{j_1}U)}{\psi}$ an amplitude
function of the particle associated with the path. Many paths yield
0 amplitude due to the decoherence projections, the $A_{xj}$'s.
However, the summation in (\ref{defdq}) over all paths $(j_1, j_2,
\ldots j_t, (x,n))$ gives the probability of observing the particle
at position state $\ket{x}$ at time $t$.

At each step of a path, the walk is either not measured with
probability $q=1-p$ or is measured at $\ket{x,n}$ with probability
$p$. So when $p=0$, the walk is not measured and the system is the
same as the pure quantum random walk previously defined. When $p=1$,
the particle is interfered with at each step, hence the quantum
behavior essentially disappears and the system is exactly classical.

We work on the decoherent Hadamard walk starting from position $0$.
We use
\begin{align}
P_t({\ket{\psi}},\ket{\phi})= \sum_{j_t\in \mathbb{A}} \ldots
\sum_{j_1\in \mathbb{A}} | \lo{\phi}{
(A_{j_t}U)\ldots(A_{j_1}U)}{\psi}|^2
\end{align}
to denote the probability that, at time $t$, of a particle in the
decoherent quantum random walk starting from $\ket{\psi}$ to be
found at state $\ket{\phi}$. In particular, we denote the
probability that at time $t$, the particle starting at $\ket{0,m}$
can be found at $\ket{x,n}$ by
\begin{equation}
\begin{split}
 & P_{m,n}(x,t)  \\
=& \sum_{j_1, \ldots, j_t \in \mathbb{A}}|\lo{x,n}{
(A_{j_t}U)\ldots(A_{j_1}U)}{0,m}|^2.
\end{split}
\end{equation}

Since we are interested in the limiting distribution of the walk, we
focus on the Fourier transform of the above probabilities,
\begin{align}
\widehat{P}_{m,n}(k,t)=\sum_x P_{m,n} (x,t) e^{i k x}.
\end{align}

We consider two types of walks. We first consider the walk starting
at the state $\ket{\phi_0}={1 \over \sqrt{2}}\ket{0,1} + i {1 \over
\sqrt{2}}\ket{0,2}$. We call this walk ``symmetric'' and denote it
by $X_t$. Note that the characteristic function of $X_t$ is
\begin{equation} \label{ct}
\begin{split}
\widehat{P}(k,t)= {1 \over 2} \sum_m \sum_n \widehat{P}_{m,n}(k,t).
\end{split}
\end{equation}
From the above equation, we can see that its characteristic function
is obtained by taking the average of those with initial chirality
state $m$. Furthermore, in \cite{AM01}, it is shown that the pure
quantum random walk starting with $\ket{\phi_0}$ has a symmetric
position distribution. These are the reasons why we call it
``symmetric.''

We also consider the walk that starts at $\ket{0,1}$, $\tilde{X_t}$,
i.e, the walk starting at 0 with chirality ``right.'' In this case,
the characteristic function of $\tilde{X_t}$ is
\begin{align}
\widehat{\tilde{P}}(k,t)= \sum_n \widehat{P}_{1,n} (k,t).
\end{align}

Our goals are to show that as $t \rightarrow \infty$,
\begin{equation}
\widehat{P}(k,t) \rightarrow e^{-{1 \over 2} v k^2},
\end{equation}
for some positive number $v$ in the symmetric walk case, as well as
to show that as $t \rightarrow \infty$,
\begin{equation}
\widehat{\tilde{P}}(k,t) \rightarrow e^{-{1 \over 2} v k^2}
\end{equation}
in this specific initial state case.

\section{Generating Functions and the Decoherence Equation}

\subsection{Generating Functions}
The direct calculation involves some formidable, very complicated
combinatorics. Therefore, we introduce the idea of generating
functions. The \emph{generating function} of the decoherent quantum
random walk is
\begin{align}
P_{m,n}(x,z)=\sum_{t=0}^{\infty} P_{m,n}(x,t) z^t.
\end{align}
The Fourier transform of the generating function is
\begin{align}
\widehat{P}_{m,n}(k,z)=\sum_{x} P_{m,n}(x,z)e^{i k x}.
\end{align}
Note that for $z$ in the unit disk $\{z: |z|<1 \}$, since
$|\widehat{P}_{m,n}(k,t)| \leq 1$ and $|P_{m,n}(x,t)| \leq 1$ for
every $t$, the $\sum_{t=0}^{\infty} \widehat{P}_{m,n}(k,t) z^t$'s
and $P_{m,n}(x,z)$'s are analytic. Furthermore,
\begin{align}
\sum_x \sum_{t=0}^{\infty}| P_{m,n}(x,t)e^{i k x} z^t|^2 < \infty.
\end{align}
Hence, by Fubini's theorem, we have
\begin{equation}
\widehat{P}_{m,n}(k,z)= \sum_{t=0}^{\infty} \widehat{P}_{m,n}(k,t)
z^t,
\end{equation}
i.e., $\widehat{P}_{m,n}(k,z)$'s are analytic and
$\widehat{P}_{m,n}(k,t)$'s are the coefficients of $z^t$ in the
expansions of $\widehat{P}_{m,n}(k,z)$'s. Therefore, instead of
finding $\widehat{P}_{m,n}(k,t)$'s directly, we first find the
explicit formulae of $\widehat{P}_{m,n}(k,z)$'s and then we apply
Cauchy's Theorem
\begin{align}
\widehat{P}_{m,n}(k,t)={1 \over 2 \pi i} \oint_{|z|=r}
{\widehat{P}_{m,n}(k,z) \over z^{t+1}} dz,
\end{align}
for some $0<r<1$, to get $\widehat{P}_{m,n}(k,t)$.

\subsection{The Decoherence Equation}
The functions $\widehat{Q}_{m,n}(k,z)$ and $Q_{m,n}(x,z)$ are very
important in our proofs. We let $\widehat{W}_{m,n}(k,t)=\sum_x
W_{m,n}(x,t) e^{ikx}$ be the Fourier transform of the pure Hadamard
walk. We also let $\widehat{Q}_{m,n}(k,z)={p \over q}
\sum^{\infty}_{t=1} \widehat{W}_{m,n}(k,t) (qz)^t$ for $0 <p \leq 1$
and $q=1-p$. Note that $|\widehat{W}_{m,n}(k,t)| \leq 1$. Hence, for
$z \in \{z:|z|<{1 \over q}\}$, $|\widehat{Q}_{m,n}(k,z)| < \infty$.
Therefore, $\widehat{Q}_{m,n}(k,z)$ is analytic in $\{z:|z|<{1 \over
q}\}$. Furthermore, let $Q_{m,n}(x,z)={p \over q}
\sum^{\infty}_{t=1} W_{m,n}(x,t)(qz)^t$, and by Fubini's theorem
again we have
\begin{align}\label{qeqn}
\widehat{Q}_{m,n}(k,z)={p \over q} \sum^{\infty}_{t=1}
\widehat{W}_{m,n}(k,t) (qz)^t = \sum_x Q_{m,n}(x,z) e^{i k x}.
\end{align}

Using the above notations, we derive the following theorem, whose
proof could be found in the appendix and in \cite{ZH07}.

\begin{tem}{(The Decoherence Equation)}\label{thm1}\\
The functions $\widehat{P}_{m,n}(k,z)$'s are analytic in $\{z:
|z|<1\}$ and are meromorphic in $\{z: |z| < {1 \over q}\}$.
Furthermore, if we denote the matrices of $(\widehat{P}_{m,n}(k,z))$
and $(\widehat{Q}_{m,n}(k,z))$ by $P$ and $Q$ respectively, then
\begin{align}\label{decoeqn}
P = -{q \over p} I+ {1 \over p}{({I - Q})}^{-1}.
\end{align}
\end{tem}

This equation establishes the relationship between the decoherent
quantum random walk (left-hand side) and the pure quantum random
walk (right-hand side). By working on the Fourier transform of the
pure quantum random walk, we get the formulae of
$\widehat{P}_{m,n}(k,z)$'s from this equation.

\section{Main Results}
We list our results for the two types of decoherent quantum random
walks here. The step-by-step mathematical proofs of theorems in this
section could be found in \cite{ZH07}.

\subsection{Results for the Symmetric Decoherent Quantum Random Walk}\label{sit:1}
The following theorem gives the closed form formula of the Fourier
transform of the generating function of the walk. This formula
synthesizes all the information of the walk and is crucial in
proving our results.

\begin{tem}\label{fgf}
The Fourier transform of the generating function of the symmetric
decoherent Hadamard walk, $\widehat{P}(k,z)$, is given by
\begin{equation}\label{fgformula}
\begin{split}
 & \widehat{P}(k,z)\\
=& {q (q - \cos^2{k}) z^2 + p \cos{k} z + (1 - z \cos{k}) E \over p
q \cos{k} z^3 - (p q + p) z^2 + p \cos{k} z + (z^2 - 2 \cos{k} z +
1) E }
\end{split}
\end{equation}
where
\begin{equation}\!E = \sqrt{(q^2z^2 - (1 + \cos{k}) q z
+1)(q^2z^2 + (1 + \cos{k}) q z +1)}
\end{equation}
 and the square root in the
formula of $E$ is defined through the principal logarithm.
\end{tem}

We first show that the position distribution of the walk is
symmetric with respect to the origin.

\begin{tem}\label{thm2.3}
Let $X_t$ be the symmetric decoherent Hadamard walk on the line with
$0 < p \leq 1$ and $q = 1 - p$. Then $E(X_t)=0, \forall t$.
\end{tem}

We then consider the limiting distribution. We derive the following
theorem for the limiting distribution of the symmetric decoherent
Hadamard walk.

\begin{tem}\label{thm2.4}
Let $X_t$ be the symmetric decoherent Hadamard walk on the line with
$0 < p \leq 1$ and $q = 1 - p$. Then the characteristic function
$\widehat{P}(k,t)$ of $X_t$ satisfies
\begin{align}
\widehat{P}({k \over \sqrt{t}},t)= \exp{(-{p+2\sqrt{1+q^2}-2 \over
2p} k^2)}+O(t^{-1})
\end{align}
as $t \rightarrow \infty$, i.e.,
\begin{align}
{X_{t} \over \sqrt{t}} \rightarrow N(0, {p+2\sqrt{1+q^2}-2 \over p})
\end{align}
in distribution as $t \rightarrow \infty$.
\end{tem}

This theorem states that after a long time, the position
distribution induced by the wave function of the particle is
Gaussian. We see from the variance of the distribution that it is a
mixture of the quantum and classical limiting distribution.

We give a plot of different limiting distributions w.r.t. different
$p$'s in Figure 1.
\begin{center}
\epsfig{file=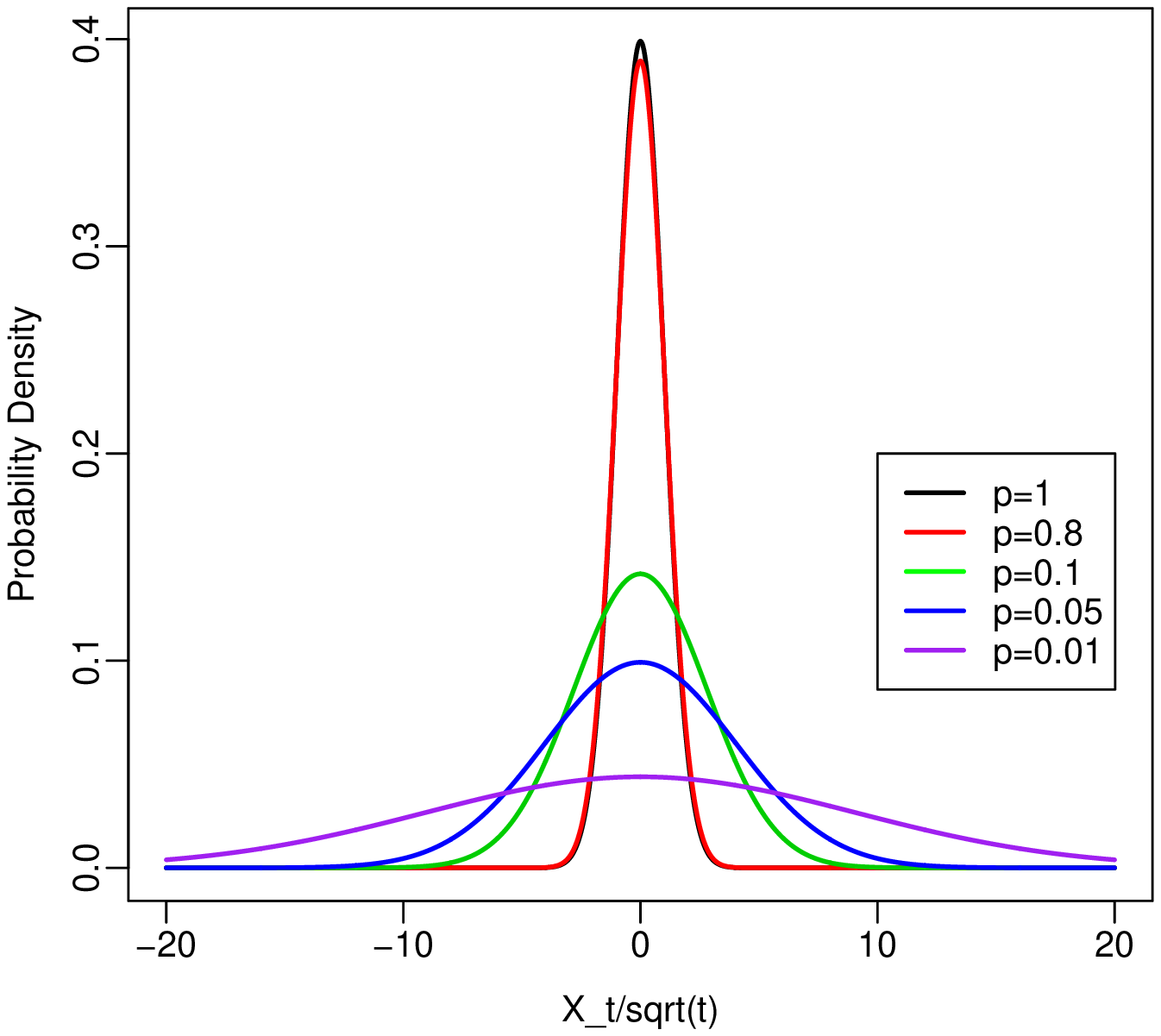,height=3.2in}\\
\small Figure 1. Limiting position distributions of ${X_t \over
\sqrt{t}}$ when p=1,0.8,0.1,0.05 and 0.01.
\end{center}

We see that when $p$ increases to 1, the distribution is
asymptotically standard normal, as in a classical walk. On the other
hand, when $p$ is decreasing to 0, the variance increases to
infinity, meaning that ${X_t \over \sqrt{t}}$ does not converge.
\\
\\
We also find the long-term variance of $X_t$ as follows.

\begin{tem}\label{thm2.5}
Let $X_t$ be the symmetric decoherent Hadamard walk on the line with
$0 < p \leq 1$ and $q = 1 - p$. Then
\begin{equation}\label{decovar1}
\begin{split}
 & Var(X_t) \\
=&{p+2\sqrt{1+q^2}-2 \over p} t - {2 q^2 \over p \sqrt{1+q^2}} \\
 &-{2 \over p^2} (1+q^2-\sqrt{1+q^2}) + O(e^{- c t}),
\end{split}
\end{equation}
 for some $c>0$, as $t \rightarrow \infty$.
\end{tem}

This theorem shows that for fixed $p$ and large $t$, the magnitude
of the walk is growing linearly in $\sqrt{t}$.

\begin{center}
\epsfig{file=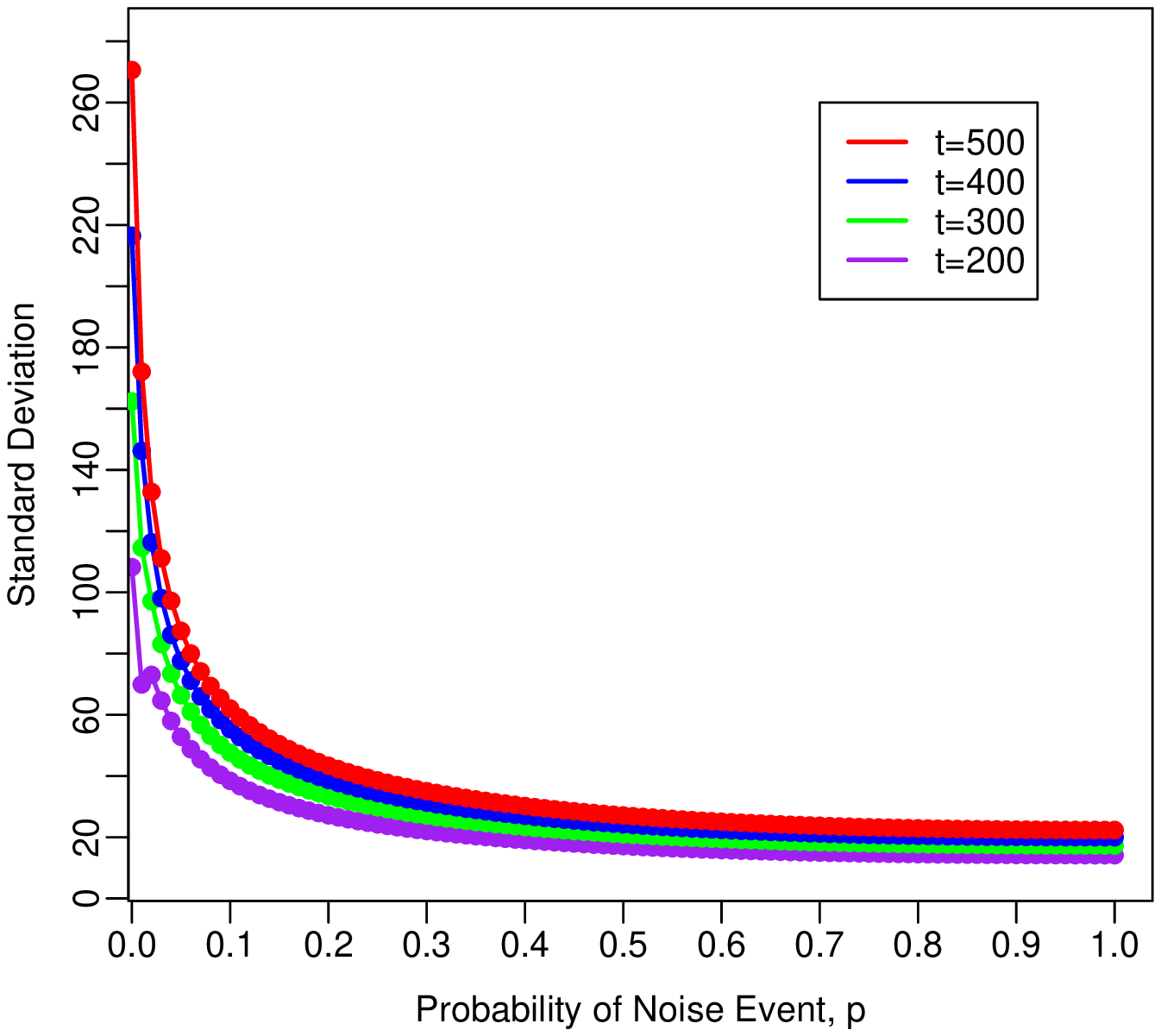,height=3.2in}\\
\small Figure 2. Standard deviation of the particle position as a
function of $p$ at t=200, 300, 400 and 500. The grid of $p$ is from
0 to 1 with increment 0.01.
\end{center}

We plotted the standard deviations obtained from (\ref{decovar1})
against $p$'s in Figure 2. This picture compares very well with
Figure 1 in \cite{KEN02}, where the values of standard deviations
are from numerical simulations. Note that when $t=200$ and $p=0.01$,
the standard deviation seems too low. This is because when $t=200$,
the limiting phenomenon has not occurred yet for the walk with
$p=0.01$. Therefore, the formula (\ref{decovar1}) is not a good
approximation and the decoherent walk is still similar as
``quantum''. We discuss this in Section \ref{pseudo}.

\subsection{Results for the Decoherent Hadamard Walk Starting at $\ket{0,1}$.}\label{sit:2}
Now we consider the decoherent walk starting at $\ket{0,1}$. As
before we first find its expectation.
\begin{tem}\label{thm2.6}
Let $\tilde{X_t}$ be the decoherent Hadamard walk starting at
$\ket{0,1}$ with $0 < p \leq 1$ and $q = 1 - p$. Let
$\mu_t=E(\tilde{X_t})$, then we have $\mu_t={\sqrt{1+q^2}-1 \over
p}+O(e^{-d t})$ for some $d>0$, as $t \rightarrow \infty$.
\end{tem}

This theorem shows that the limiting expected position of the
decoherent Hadamard walk is to the right of the origin, if the
initial coin state is ``right.'' We see that when $p \rightarrow 0$,
$\mu_t \rightarrow \infty$. It is consistent with the result in
\cite{AM01} that the pure quantum random walk starting with
chirality ``right'' drifts to the right.

For the second moment, we have the same result as for the symmetric
walk.

\begin{tem}\label{thm2.7}
 Let $\tilde{X_t}$ be the decoherent Hadamard walk starting at
$\ket{0,1}$ with $0 < p \leq 1$ and $q = 1 - p$. Then
\begin{equation}
\begin{split}
 & Var(\tilde{X_t}) \\
=&{p+2\sqrt{1+q^2}-2 \over p} t - {2 q^2 \over p \sqrt{1+q^2}}\\
 &-{2 \over p^2} (1+q^2-\sqrt{1+q^2}) + O(e^{- c t}),
\end{split}
\end{equation}
 for some $c>0$, as $t \rightarrow \infty$.
\end{tem}

Now we show that the limiting position distribution of the
decoherent Hadamard walk starting at $\ket{0,1}$ is also Gaussian.
\begin{tem}\label{thm2.8}
Let $\tilde{X_t}$ be the decoherent Hadamard walk starting at
$\ket{0,1}$ with $0 < p \leq 1$ and $q = 1 - p$. Then the
characteristic function $\widehat{\tilde{P}}(k,t)$ of $\tilde{X_t}$
satisfies
\begin{align}
\widehat{\tilde{P}}({k \over \sqrt{t}},t)= \exp{(
-{p+2\sqrt{1+q^2}-2 \over 2p} k^2)}+O(t^{-{1 \over 2}})
\end{align}
as $t \rightarrow \infty$, i.e.,
\begin{align}
{\tilde{X_{t}} - \mu_t \over \sqrt{t}} \rightarrow N(0,
{p+2\sqrt{1+q^2}-2 \over p})
\end{align}
in distribution as $t \rightarrow \infty$.
\end{tem}

\begin{rem}
Note that here the converging speed is $O(t^{-{1 \over 2}})$ while
we have $O(t^{-1})$ for the symmetric walk. This is because when one
takes the average of the $\widehat{P}_{m,n}({k \over
\sqrt{t}},t)$'s, the error terms in $O(t^{-{1 \over 2}})$ cancel out
one another. This result shows that the symmetric walk converges
faster.
\end{rem}

\section{Speed of the Walk When $p$ is Small: Pseudoquantum
Phenomenon}\label{pseudo} In \cite{AM01}, it is shown that the
long-term variance of the Hadamard walk is $(1 - {1 \over
\sqrt{2}})t^2$. We proved the same result via our approach by
letting $p \rightarrow 0$.
\begin{tem}\label{pqv}
Let $Q_t$ denote the Hadamard walk on the line, the long-term
variance satisfies ${Var(Q_t) \over t^2} \rightarrow (1 - {1 \over
\sqrt{2}})$ as $t \rightarrow \infty$.
\end{tem}
We also investigated the case when $p$ is small. Because current
literature shows that the distribution of the pure Hadamard walk
compares well with the uniform distribution over $[-{t \over
\sqrt{2}}, {t \over \sqrt{2}}]$, we also compare the decoherent
Hadamard walk with it. Denote the uniform distribution over $[-{t
\over \sqrt{2}}, {t \over \sqrt{2}}]$ by $U_t$. We shall compare the
variance of $U_t$ and the variance of the symmetric walk $X_t$. Note
that by (\ref{decovar1}) and
\begin{equation}
Var(U_t) = {t^2 \over 6}.
\end{equation}
The difference $Var(U_t)-Var(X_t)$ is minimized at
\begin{align}\label{diffvar}
t_0={6 (\sqrt{1+q^2}-1) \over p} + 3.
\end{align}

The minimizer $p$ from (\ref{diffvar}) when $t_0=200$ is about
$0.0014$. This compares well with the numerical result in
\cite{KEN02}, which demonstrated that when $t_0=200$, the $p$ that
minimizes the difference between $X_t$ and $U_t$ is about $0.0013$.
A plot of $t_0$ versus $p$ is also given in Figure 3.
\begin{center}
\epsfig{file=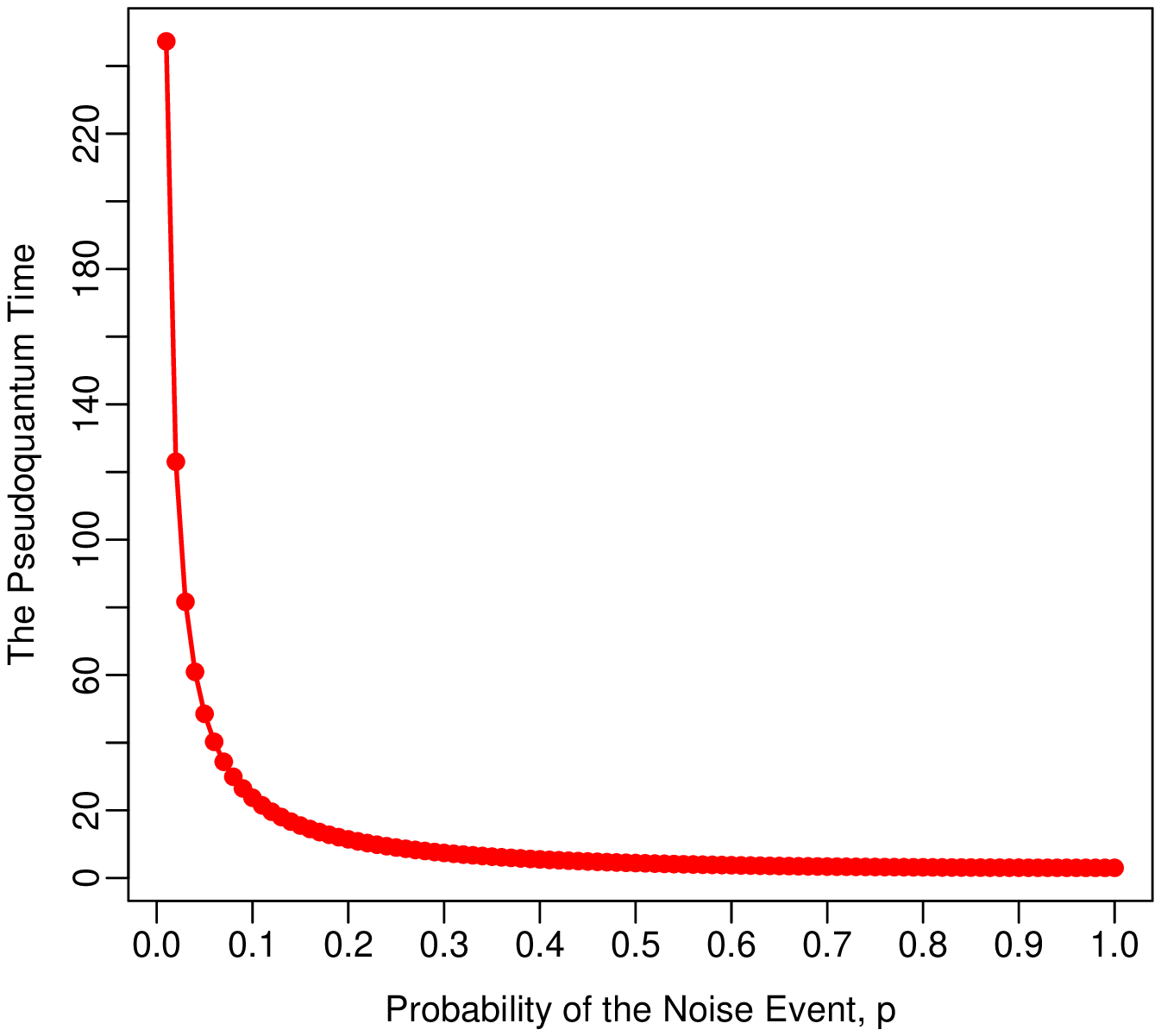,height=3.2in}\\
\small Figure 3. The pseudoquantum time $t_0$ as a function of $p$.
The grid of $p$ is from 0.01 to 1 with increment 0.01.
\end{center}

The time $t_0$ is interesting in the sense that the variance of the
decoherent walk could not be regarded as linear in $t$ before $t_0$,
i.e, the walk before $t_0$ could not be regarded as ``classical.''
We call the period from $0$ to $t_0$ ``pseudoquantum'' since the
walk takes on quantum features. After $t_0$, the variance of the
walk approaches the formula obtained in (\ref{decovar1}). For
example, for $p=0.01$, $t_0=247.3$. Therefore, when $t=200$, the
limiting behavior has not occurred and the decoherent Hadamard walk
has more quantum features than classical ones.

\section{Conclusions} We have investigated the quantum walk with
decoherence on both position and chirality states. Long-term limits
are obtained for both the symmetric walk and the walk starting at 0
with chirality ``right.'' We provide analytical explanations of the
dynamics of the decoherent quantum walk system and we see that the
system is indeed a mixture of quantum and classical ones. The
limiting distributions of quantum random walks are shown to be
Gaussian if decoherence occurs. These results are very important
properties of the decoherent quantum random walks and could be
essential for the development of quantum algorithms and experiments.

We also see that when $p$ is small, the system remains non-classical
for a very long time. If a quantum algorithm can be finished before
the classical features appear, then we may call it a
``pseudoquantum'' algorithm. However, we do not know how fast the
``pseudoquantum'' algorithms are as compared to the classical ones.
Therefore, we suggest future studies on these areas.

\section{Acknowledgments}
This paper is based on the Ph.D.
dissertation of Kai Zhang. Kai is grateful to Professor Wei-Shih
Yang for his excellent role as the thesis advisor. He provided
encouragement, sound advice and great ideas throughout the
thesis-writing period. In particular, the idea of the decoherence
equation is suggested by Professor Yang. Kai appreciates Professor
Shiferaw Berhanu, Professors Janos Galambos, Professor Seymour
Lipschutz and Professor Yuan Shi for their insightful advice too.
Kai would also like to thank Professor Boris Datskovsky, Professor
Gonzalo Abal, Professor Andris Ambainis, Professor Todd Brun,
Professor Raul Donangelo and Professor Viv Kendon for helpful
communications.

\appendix
\section{Proof of Theorem \ref{thm1}.}
We start with an observation about the decoherent quantum random
walk and get a recursive formula. Then we apply that formula to the
$\widehat{P}_{m,n}(k,z)$'s to establish the decoherence equation.

For any state $\ket{\phi} \in H$, $\ket{\phi}$ can be written as
$\ket{\phi}=\sum_{y,l} \ip{y,l}{\phi} \ket{y,l}.$ By definition,
\begin{align}
\begin{split}
& P_{t+1} (\ket{0,m},\ket{\phi})\\
=&q P_t (\ket{0,m}, U^* \ket{\phi}) + p \sum_{y,l}
|\ip{y,l}{\phi}|^2 P_t (\ket{0,m},U^* \ket{y,l}).
\end{split}
\end{align} In particular, for $\ket{\phi}=\ket{x,n}$, we have
\begin{align}
 P_{t+1} (\ket{0,m},\ket{\phi_{x,n}})=  P_{t} (\ket{0,m}, U^* \ket{x,n}),
\end{align}
which in turn gives
\begin{align}\label{recur1}
\begin{split}
& P_{t+1} (\ket{0,m}, \ket{\phi})\\
= &q P_t (\ket{0,m}, U^* \ket{\phi}) + p \sum_{y,l}
|\ip{y,l}{\phi}|^2 P_{t+1} (\ket{0,m}, \ket{y,l}).
\end{split}
\end{align}
This is our recursive formula. Also, for $t=1$, we have
\begin{align}\label{recur2}
\begin{split}  &P_1 (\ket{0,m},\ket{\phi})\\
= & q |\lo{\phi}{ U }{0,m}|^2 + p \sum_{y,l} |\ip{y,l}{ \phi}|^2 P_1
(\ket{0,m}, U^* \ket{y,l})
\end{split},
\end{align}
and
\begin{align}
P_1 (\ket{0,m},\ket{x,n})=|\lo{x,n}{U}{0,m}|^2 .
\end{align}
Apply the recursive formula (\ref{recur1}) and (\ref{recur2})
repeatedly, we have the following equation
\begin{equation}\label{recur3}
\begin{split}
 &P_{m,n}(x,t)=P_t(\ket{0,m}, \ket{x,n})\\
=&\sum_{s=1}^{t-1}p q^{s-1}\sum_{y,l}|\lo{y,l}
{(U^*)^s}{x,n}|^2P_{t-s}(\ket{0,m}, \ket{y,l})+ \\
&+ q^{t-1}|\lo{x,n}{U^t}{0,m}|^2.
\end{split}
\end{equation}
Note that by the definition of $W_{m,n}(x,t)$, we have that
\begin{equation}
\begin{split}
 &|\lo{y,l}{(U^*)^s}{x,n}|^2\\
=&|\lo{x,n}{U^s}{y,l}|^2\\
=&W_{l,n}(x-y,s)
\end{split}
\end{equation} and that
\begin{align}
|\lo{x,n}{U^t}{0,m}|^2=W_{m,n}(x,t).
\end{align}

Therefore, (\ref{recur3}) becomes
\begin{align}\label{recur:4}
\begin{split}
& P_{m,n}(x,t)\\
=&\sum_{s=1}^{t-1} p q^{s-1}
\sum_{y,l}W_{l,n}(x-y,s)P_{m,l}(y,t-s)+\\
& +q^{t-1}W_{m,n}(x,t).
\end{split}
\end{align}

Now, by (\ref{recur:4}), for $z \in \{z: |z|<{1 \over q}\}$,
\begin{equation}\label{recur:5}
\begin{split}
& P_{m,n}(x,z)\\
=& \delta^{0,m}_{x,n} + {1 \over p} Q_{m,n}(x,z) -
           Q_{m,n}(x,z)+\\
 & +\sum_{y,l} Q_{l,n}(x-y,z)P_{m,l}(y,z),
\end{split}
\end{equation}
where
\begin{displaymath}
\delta^{\alpha,m}_{\beta,n}= \left \{
\begin{array}{ll}
1 ,& \alpha=\beta, m=n\\
0 ,& otherwise
\end{array}
\right. .
\end{displaymath}

Finally, we take the Fourier transform on (\ref{recur:5}) to get
\begin{equation}\label{recur:6}
\begin{split}
& \widehat{P}_{m,n}(k,z)\\
=& \delta^m_n + {q \over p} \widehat{Q}_{m,n}(k,z) + \sum_l
\widehat{P}_{m,l}(k,z)\widehat{Q}_{l,n}(k,z),
\end{split}
\end{equation}
where
\begin{displaymath}
\delta^m_n= \left \{
\begin{array}{ll}
1 ,& m=n\\
0 ,& otherwise
\end{array}
\right. .
\end{displaymath}
The interchanges of summation are justified since the series
absolutely converges. Now, denoting the matrices
$(\widehat{P}_{m,n}(k,z))$ and $(\widehat{Q}_{m,n}(k,z))$ by $P$ and
$Q$, we have the following equation:
\begin{align}
P = I + {q \over p} Q + PQ,
\end{align}
which is equivalent to
\begin{align}\label{deco2}
P(I -Q) = - {q \over p}(I - Q) + {1 \over p} I.
\end{align}

We complete the proof by the following lemma.

\begin{lem}\label{inv}
For $z \in \{z: |z|< 1 \}$, the matrix $I-Q$ is invertible.
\end{lem}
\emph{Proof.}
 For $z \in \{z: |z|< 1 \}$, if we let $Q_{m,n}=\widehat{Q}_{m,n}(k,z)$, we have
\begin{equation} \label{qst}
\begin{split}
 &|Q_{m,1}| + |Q_{m,2}|\\
 =& {p \over q} \sum_{t=1}^{\infty}
|\widehat{W}_{m,1}(k,t)(q
z)^t| + {p \over q} \sum_{t=1}^{\infty}|\widehat{W}_{m,2}(k,t) (q z)^t|\\
<& {p \over q} \sum_{t=1}^{\infty} q^t (|\widehat{W}_{m,1}(k,t)|+|\widehat{W}_{m,2}(k,t)|)\\
\leq& {p \over q} \sum_{t=1}^{\infty} q^t (|\widehat{W}_{m,1}(0,t)|+|\widehat{W}_{m,2}(0,t)|)\\
\leq& {p \over q} {q \over p}=1, \forall i .
\end{split}
\end{equation}

(\ref{qst}) implies that $\|Q\|_{\infty} = \max_m \sum_n |Q_{m,n}|<
1$. Therefore,

\begin{equation}
\|\sum_{j=0}^{\infty} Q^j\|_{\infty}     \leq \sum_{j=0}^{\infty}
\|Q^j\|_{\infty}  <  \infty,
\end{equation}
i.e., the series $\sum_{j=0}^{\infty} Q^j$ converges. This implies
that $(I-Q)^{-1}$ exists and $$(I-Q)^{-1}=\sum_{j=0}^{\infty} Q^j.$$

By Lemma \ref{inv}, $I-Q$ is invertible and together with
(\ref{deco2}) we have
\begin{align}
P=-{q \over p} I + {1 \over p} (I-Q)^{-1},
\end{align}
which is exactly (\ref{decoeqn}).

For $z \in \{z: |z|< {1 \over q} \}$, $|\det(I - Q)| < \infty$.
Hence, $\det{(I-Q)}$ is analytic. Note also that
\begin{equation}
\begin{split}
\widehat{P}_{1,1}(k,z) =& - {q \over p} + {1-Q_{2,2} \over
p \det{(I-Q)}},\\
\widehat{P}_{1,2}(k,z) =& {Q_{1,2} \over p \det{(I-Q)}},\\
\widehat{P}_{2,1}(k,z) =& {Q_{2,1} \over p \det{(I-Q)}},\\
\widehat{P}_{2,2}(k,z) =& - {q \over p} +  {1-Q_{1,1} \over p
\det{(I-Q)}}.
\end{split}
\end{equation}
Therefore, $\widehat{P}_{m,n}(k,z)$'s are meromorphic functions for
$z \in \{z: |z|< {1 \over q} \}$.

\section{Proof of Theorem \ref{fgf}.}
To get formula (\ref{fgformula}), first we need to know the formulae
of $\widehat{W}_{m,n}(k,t)$, i.e., we look at the pure quantum walk
in the Fourier transform.

Similar to the setup in \cite{AM01}, if we let the initial state be
$\ket{0,m}$, and we let $\Psi_{m,n}(x,t)=\lo{x,n}{ U^t}{0,m}$ be the
coefficient of the walk at time $t$ at coordinate $\ket{x,n}$, then
$W_{m,n}(x,t)=|\Psi_{m,n}(x,t)|^2$. We also introduce
$\widehat{\Psi}_{m,n}(k,t)=\sum_x \Psi_{m,n}(x,t) e^{i k x}$ and
$\widehat{\Psi}_m(k,t)=(\widehat{\Psi}_{m,1}(k,t),
\widehat{\Psi}_{m,2}(k,t))^T $ in the Fourier transform as in
\cite{AM01}. The evolution operator in Fourier transform space,
$U(k)$, is defined s.t.
$\widehat{\Psi}_m(k,t+1)=U(k)\widehat{\Psi}_m(k,t)$. It is obtained
in \cite{AM01} that
\begin{equation}
U(k)= \frac{1}{\sqrt{2}} \left (
\begin{array}{ll}
e^{ik} & e^{ik}\\
e^{-ik} & -e^{-ik}
\end{array}
\right ).
\end{equation}

Therefore, if we let $A_k={1 \over 2}+{\cos{k} \over 2
\sqrt{1+\cos^2{k}}}$ and $C_k={e^{-ik} \over 2 \sqrt{1+\cos^2{k}}}$,
then for $t=2j-1$, we have
\begin{equation}
\begin{split}
& U^t(k) \\
=& \left (
\begin{array}{ll}
-e^{-i \omega_k t} + 2 A_k \cos{\omega_k t} & 2
\bar{C}_k\cos{\omega_k t} \\
2 C_k \cos{\omega_k t} & e^{i \omega_k t} - 2 A_k \cos{\omega_k t}
\end{array}
\right ). \end{split}
\end{equation} Also, for $t=2 j$, we have
\begin{equation}
\begin{split}
& U^t(k) \\
=& \left (
\begin{array}{ll}
e^{-i \omega_k t} + 2 A_k i \sin{\omega_k t} & 2
i \bar{C}_k\sin{\omega_k t} \\
2 i C_k \sin{\omega_k t} & e^{i \omega_k t} - 2 A_k i \sin{\omega_k
t}
\end{array}
\right ). \end{split}
\end{equation}

Now, note that
\begin{align}
\widehat{\Psi}_{m,n}(k,0)=\sum_x\ip{x,n}{0,m} e^{i k x}=\delta^m_n,
\end{align}
and that $\widehat{\Psi}_m(k,t)=(U(k))^t \widehat{\Psi}_m(k,0)$, we
conclude that $\widehat{\Psi}_{m,n}(k,t)=(U^t(k))_{n,m}$.

Hence, for $t=2j -1$,
\begin{equation}
\begin{split}
\widehat{\Psi}_{1,1}(k,t)=&-e^{-i \omega_k t}+2 A_k \cos{\omega_k t},\\
\widehat{\Psi}_{1,2}(k,t)=&2C_k \cos{\omega_k t},\\
\widehat{\Psi}_{2,1}(k,t)=&2\bar{C}_k \cos{\omega_k t},\\
\widehat{\Psi}_{2,2}(k,t)=&e^{i\omega_k t}-2A_k \cos{\omega_k t}.
\end{split}
\end{equation} For $t=2 j$,
\begin{equation}
\begin{split}
\widehat{\Psi}_{1,1}(k,t)=&e^{-i \omega_k t}+2 A_k i \sin{\omega_k t},\\
\widehat{\Psi}_{1,2}(k,t)=&2 i C_k \sin{\omega_k t},\\
\widehat{\Psi}_{2,1}(k,t)=&2 i \bar{C}_k \sin{\omega_k t},\\
\widehat{\Psi}_{2,2}(k,t)=&e^{i \omega_k t}-2 A_k i \sin{\omega_k
t}.
\end{split}
\end{equation}

Since $W_{m,n}(x,t)=|\Psi_{m,n}(x,t)|^2$, in the Fourier transform,
\begin{align}
\widehat{W}_{m,n}(k,t)= {1 \over 2 \pi} \int_0^{2 \pi}
\widehat{\Psi}_{m,n}(s,t)\widehat{\Psi}_{m,n}(k-s,t) d s.
\end{align}

We separate the real and imaginary parts of $\widehat{W}_{m,n}(k,t)$
and get their formulae as follows. For $t=2j -1$,
\begin{align}
&  Re( \widehat{W}_{1,1}(k,t))=Re( \widehat{W}_{2,2}(k,t))\\
&= {1 \over 2 \pi} \int^{2 \pi}_0 {1 \over 2}{\cos{\omega_s t}
\cos{\omega_{k-s} t} \over \cos{\omega_s} \cos{\omega_{k-s}}}
\cos{s}\cos{(k-s)} d s -\nonumber \\
 & -{1 \over 2 \pi}\int^{2 \pi}_0 \sin{\omega_s t}\sin{\omega_{k-s}
t} d
 s,\nonumber\\
& Re( \widehat{W}_{1,2}(k,t))=Re(\widehat{W}_{2,1}(k,t)) \\
&= {1 \over 2 \pi} \int^{2 \pi}_0 {\cos{k} \over 2} { \cos{\omega_s
t}\cos{\omega_{k-s} t}\over
\cos{\omega_s} \cos{\omega_{k-s}}}  d s,\nonumber\\
&  Im(\widehat{W}_{1,1}(k,t))=-Im(\widehat{W}_{2,2}(k,t))\\
&= {1 \over 2 \pi} \int^{2 \pi}_0 ( {1 \over \sqrt{2}} {\cos{s}
\over \cos{\omega_s}} \cos{\omega_s t} \sin{\omega_{k-s} t}+\nonumber\\
 &+{1 \over \sqrt{2}} {\cos{(k-s)} \over \cos{\omega_{k-s}}}
\cos{\omega_{k-s}
t} \sin{\omega_s t} ) d s,\nonumber\\
 & Im(\widehat{W}_{1,2}(k,t))=-Im(\widehat{W}_{2,1}(k,t))\\
&=-{1 \over 2 \pi} \int^{2 \pi}_0 {\sin{k} \over 2}{\cos{\omega_s t}
\cos{\omega_{k-s} t} \over \cos{\omega_s} \cos{\omega_{k-s}}} d s.
\nonumber
\end{align}
For $t=2j$,
\begin{align}
&  Re(\widehat{W}_{1,1}(k,t))=Re( \widehat{W}_{2,2}(k,t)) \\
&= -{1 \over 2 \pi} \int^{2 \pi}_0 {1 \over 2}{\sin{\omega_s t}
\sin{\omega_{k-s} t} \over \cos{\omega_s} \cos{\omega_{k-s}}}
\cos{s}\cos{(k-s)} d s +\nonumber\\
& + {1 \over 2 \pi}\int^{2 \pi}_0 \cos{\omega_s t}\cos{\omega_{k-s} t} d s,\nonumber\\
\nonumber\\
&  Re( \widehat{W}_{1,2}(k,t))=Re(\widehat{W}_{2,1}(k,t))\\
&= -{1 \over 2 \pi} \int^{2 \pi}_0 {\cos{k} \over 2} {\sin{\omega_s
t}\sin{\omega_{k-s} t} \over
\cos{\omega_s} \cos{\omega_{k-s}}}  d s,\nonumber\\
\nonumber\\
&  Im(\widehat{W}_{1,1}(k,t))=-Im(\widehat{W}_{2,2}(k,t))\\
 &= {1
\over 2 \pi} \int^{2 \pi}_0 ( {1 \over \sqrt{2}} {\cos{s} \over
\cos{\omega_s}} \sin{\omega_s t} \cos{\omega_{k-s} t}+\nonumber\\ &
+{1 \over \sqrt{2}} {\cos{(k-s)} \over \cos{\omega_{k-s}}}
\sin{\omega_{k-s} t} \cos{\omega_s t} ) d s,\nonumber\\
\nonumber\\
&  Im(\widehat{W}_{1,2}(k,t))=-Im(\widehat{W}_{2,1}(k,t))\\
&={1 \over 2 \pi} \int^{2 \pi}_0 {\sin{k} \over 2}{\sin{\omega_s t}
\sin{\omega_{k-s} t} \over \cos{\omega_s} \cos{\omega_{k-s}}} d
s.\nonumber
\end{align}
Now we are ready to find $\widehat{P}_{m,n}(k,z)$'s formulae. We
first introduce several short notations. We introduce the
$\Sigma_i$'s for $z \in \{z: |z|< {1 \over q} \}$. Let
\begin{align}
\Sigma_1&= Re(Q_{1,1})\\
        &= {p \over q} \sum_{t=1}^{\infty} [Re(\widehat{W}_{1,1}(k,t))](q
        z)^t,\nonumber\\
\Sigma_2&= Re(Q_{1,2})\\
        &= {p \over q} \sum_{t=1}^{\infty} [Re(\widehat{W}_{2,1}(k,t))](q
        z)^t,\nonumber\\
\Sigma_3&= Im(Q_{1,1})\\
        &= {p \over q} \sum_{t=1}^{\infty} [Im(\widehat{W}_{1,1}(k,t))](q
        z)^t,\nonumber\\
\Sigma_4&= Im(Q_{1,2})\\
        &= {p \over q} \sum_{t=1}^{\infty} [Im(\widehat{W}_{2,1}(k,t))](q
        z)^t.\nonumber
\end{align}

Since $|\widehat{W}_{m,n}(k,t)| \leq 1$, for $z \in \{z: |z|< {1
\over q} \}$, the above series all converge. Therefore, $\Sigma_i$'s
are all analytic in $\{z: |z|<{1 \over q}\}$.

Now $\det(I-Q)$ can be written as
\begin{equation}
\begin{split}
 &\det{(I-Q)}\\
=&1-Q_{1,1}-Q_{2,2}+Q_{1,1}Q_{2,2}-Q_{1,2}Q_{2,1}\\
=&(1-\Sigma_1)^2-{\Sigma_2}^2+ {\Sigma_3}^2- {\Sigma_4}^2.
\end{split}
\end{equation}

Note that $\widehat{P}(k,z)={1 \over 2} \sum_{m,n}
\widehat{P}_{m,n}(k,z)$. By the decoherence equation
(\ref{decoeqn}), this function can be written as
\begin{equation}\label{psigma}
\begin{split}
 &\widehat{P}(k,z)\\
=&-{q \over p} + {1 \over 2p} {2 - Q_{1,1}+Q_{1,2}+Q_{2,1}-Q_{2,2}
\over \det{(I-Q)}}\\
=& -{q \over p} + {1 - \Sigma_1 + \Sigma_2 \over p
((1-\Sigma_1)^2-{\Sigma_2}^2+ {\Sigma_3}^2- {\Sigma_4}^2)}.
\end{split}
\end{equation}

Therefore, once we have the formulae of $\Sigma_i$'s, we have the
formulae of $\widehat{P}(k,z)$. To find $\Sigma_i$'s formulae, we
first look for the formulae for a real number $z \in (-{1 \over
q},{1 \over q})$. Then we show that they are the desired formulae
for all $z \in \{z: |z|< {1 \over q} \}$. Let
\begin{align}
I_1 &\!\!=\!\! \sum_{j=1}^{\infty} \cos{[(2 j - 1) \omega_s ]} \cos
{[(2 j
- 1)\omega_{k-s}]}(q z)^{2j-1},\\
I_2 &\!\!=\!\! \sum_{j=1}^{\infty} \sin{[(2 j - 1) \omega_s ]} \sin
{[(2 j
- 1)\omega_{k-s}]}(q z)^{2j-1},\\
I_3 &\!\!=\!\! \sum_{j=1}^{\infty} \cos{[(2 j ) \omega_s ]} \cos {[(2 j)\omega_{k-s}]}(q z)^{2j},\\
I_4 &\!\!=\!\! \sum_{j=1}^{\infty} \sin{[(2 j) \omega_s ]} \sin {[(2 j)\omega_{k-s}]}(q z)^{2j},\\
I_5 &\!\!=\!\! \sum_{j=1}^{\infty} \cos{[(2 j - 1) \omega_s ]} \sin
{[(2 j
- 1)\omega_{k-s}]}(q z)^{2j-1},\\
I_6 &\!\!=\!\! \sum_{j=1}^{\infty} \sin{[(2 j - 1) \omega_s ]} \cos
{[(2 j
- 1)\omega_{k-s}]}(q z)^{2j-1},\\
I_7 &\!\!=\!\! \sum_{j=1}^{\infty} \sin{[(2 j) \omega_s ]} \cos {[(2 j)\omega_{k-s}]}(q z)^{2j},\\
I_8 &\!\!=\!\! \sum_{j=1}^{\infty} \cos{[(2 j) \omega_s ]} \sin {[(2
j)\omega_{k-s}]}(q z)^{2j}.
\end{align}
Since the $\Sigma_i$'s are bounded, we can interchange the integral
and the summation to write the $\Sigma_i$'s as
\begin{align}
& \Sigma_1\\
=& {p \over q} {1 \over 2 \pi} \int_0^{2 \pi} ({1 \over 2} {\cos{s}
\cos{(k-s)} \over \cos{\omega_s} \cos{\omega_{k-s}}}(I_1 -
I_4)-I_2+I_3) d s,\nonumber\\
&\Sigma_2\\
=& {p \over q} {1 \over 2 \pi} \int_0^{2 \pi} {1 \over 2} {\cos{k}
\over \cos{\omega_s} \cos{\omega_{k-s}}}(I_1 -
I_4) d s,\nonumber\\
 & \Sigma_3\\
=& {p \over q} {1 \over 2 \pi} \int_0^{2 \pi} {1 \over
\sqrt{2}}[{\cos{s} \over \cos{\omega_s}}(I_5+I_7) + {\cos{(k-s)}
\over \cos{\omega_{k-s}}}(I_6 + I_8)] d s,\nonumber\\
&\Sigma_4\nonumber\\
=& {p \over q} {1 \over 2 \pi} \int_0^{2 \pi} {1 \over 2} {\sin{k}
\over \cos{\omega_s} \cos{\omega_{k-s}}}(-I_1 + I_4) d s.
\end{align}
Then we have
\begin{align}
& I_1 - I_4 \\
=&{1 \over D} \cos{\omega_s} \cos{\omega_{k-s}} q z (1 - q^2z^2),\nonumber\\
 &-I_2+I_3 \\
=& {1 \over D} [ - {1 \over 2} \sin{s} \sin{(k-s)} q z +
q^2z^2[\cos^2{s} + \cos^2{(k-s)}- \nonumber\\
 &-1]- {3 \over 2} \sin{s}
\sin{(k-s)} q^3z^3 - q^4z^4],\nonumber\\
 & I_5 + I_7\\
=& {1 \over D} {1 \over \sqrt{2}} q z \cos{\omega_s}[\sin{(k-s)} +
2 q z \sin{s} + q^2z^2 \sin{(k-s)}],\nonumber\\
 & I_6+I_8 \\
=& {1 \over D} {1 \over \sqrt{2}} q z \cos{\omega_{k-s}}[\sin{s} + 2
q z \sin{(k-s)} + q^2z^2 \sin{s}], \nonumber
\end{align}
where
\begin{equation}
\begin{split}
 &D \\
=& \cos{(k - 2 s)} (q^3 z^3 - 2 \cos{k} q^2 z^2 + q z) +\\
 &+ q^4 z^4 -\cos{k} q^3 z^3 - \cos{k} q z + 1.
\end{split}
\end{equation}
Therefore,
\begin{align}
 &¡¡\Sigma_1 \\
=& p z {1 \over 2 \pi} \int_0^{2 \pi} {1 \over D} [
\cos{(k - 2 s)} q z(\cos{k} - q z)+\nonumber\\
 & + {1 \over 2}\cos{k}+ {1 \over 2} \cos{k} q^2 z^2 - q^3
         z^3] d s,\nonumber\\
 & \Sigma_2 \\
=& {1 \over 2} p z \cos{k} (1 - q^2 z^2) {1 \over 2
\pi}\int_0^{2 \pi} {1 \over D} d s,\nonumber\\
 & \Sigma_3 \\
=& p z \sin{k} {1 \over 2 \pi} \int_0^{2 \pi} {1 \over
D}[ \cos{(k - 2 s )} q z + {1 \over 2} + {1 \over 2} q^2 z^2] d s,\nonumber\\
 & \Sigma_4\\
=& {1 \over 2} p z \sin{k} (1 - q^2 z^2) {1 \over 2 \pi}\int_0^{2
\pi} {1 \over D} d s. \nonumber
\end{align}
By the integral formula
\begin{equation}
\int {d x \over b + c \cos{a x}} = {2 \over a \sqrt{b^2-c^2}}
\arctan{(\sqrt{{b - c \over b + c} }\tan{({1 \over 2} a x)})}\\
\end{equation}
for $b > c$ and the fact that
\begin{equation}
\begin{split}
 & q^4 z^4 -\cos{k} q^3 z^3 - \cos{k} q z + 1 \\
>& q^3z^3 - 2\cos{k} q^2 z^2 + q z
\end{split}
\end{equation} for $z
\in (-{1 \over q},{1 \over q}),$ we have
\begin{equation}
\begin{split}
 & {1 \over 2 \pi} \int_0^{2 \pi} {1 \over D} d s \\
=& ((1 + q z)(1 - q z) ((q^2z^2 - (1 + \cos{k}) q z
+1)\times \\
 & \times(q^2z^2 + (1 + \cos{k}) q z +1))^{1 \over 2})^{-1}.
\end{split}
\end{equation}
 Letting
\begin{equation}
E = \sqrt{(q^2z^2 - (1 + \cos{k}) q z +1)(q^2z^2 + (1 + \cos{k}) q z
+1)},
\end{equation} we have
\begin{equation}\label{sigma}
\begin{split}
\Sigma_1 =& {p z \over q^2 z^2 - 2 \cos{k} q z +1} [\cos{k} - q z\\
          &-  {\cos{k} - 2 q z + \cos{k} q^2 z^2 \over 2 E }],\\
\Sigma_2 =& p z \cos{k}{1 \over 2 E}, \\
\Sigma_3 =& {p z \sin{k} \over q^2 z^2 - 2 \cos{k} q z +1}[1 - {1 -
q^2 z^2 \over 2 E}],\\
\Sigma_4 =& - p z \sin{k}{1 \over 2 E}.
\end{split}
\end{equation}

Now that we have obtained the formulae of $\Sigma_i$'s for $z \in
({-1 \over q} , {1 \over q})$, we can check easily by taking the
principal branch of $log$, that the formulae are analytic in $\{z:
|z|<{1 \over q}\}$. Hence, by the Analytic Continuation Theorem,
they are the desired formulae for $z \in \{z: |z|<{1 \over q}\}$.

Finally, the theorem is obtained by applying (\ref{sigma}) to
(\ref{psigma}).

\section{Other Proofs in Subsection \ref{sit:1}.}
\noindent \emph{Proof of Theorem \ref{thm2.3}}. Note that from the
formula, for some $r<1$, we have
\begin{equation}
\begin{split}
 & E(X_t) \\
=& {1 \over i} \partial_k \widehat{P}(0,t)\\
=& {1 \over 2 \pi i} \oint_{|z|=r} {\partial_k \widehat{P}(0,z)
\over i z^{ t
+ 1}} d z\\
=& 0.
\end{split}
\end{equation}
The change of the order of integration and differentiation is
justified since ${\partial_k} \widehat{P}(k,z)$ is continuous on the
contour.\\
\\
\emph{Proof of Theorem \ref{thm2.4}}. The denominator of
$\widehat{P}(k,z)$ has less than eight isolated roots. We shall now
look for the root with the smallest absolute value. This root has no
closed form. However, since we concentrate on the asymptotic
behaviors, we need only to know its behaviors around $k=0$. The
properties of this root are summarized in the following lemma.
\begin{lem}
Let $D(k,z)$ denote the denominator of $\widehat{P}(k,z)$. Then the
root of $D(k,z)=0$ in $z$, with $z=1$ when $k=0$ is of the smallest
absolute value in a neighborhood of $k=0$. If we denote it by
$z(k)$, then $z(k)$ has multiplicity one and can be written as
follows.
\begin{equation}
z(k) = 1 + \partial_k z(0) k + o(k).
\end{equation}
\end{lem}
\emph{Proof.} For $k=0$, $D(0,z)=(1-z)(1-qz)(p z +(1 -
z)\sqrt{1+q^2z^2})$. By solving this equation we can see that $z=1$
has the smallest absolute value. The root of the second smallest
absolute value has a closed form expression, which can be found in
the appendix of \cite{ZH07}. We denote this root by $\tilde{z}(p)$.
An expansion of the root around $p=0$ is
\begin{align}\label{zp}
\tilde{z}(p)= 1 + {\sqrt{2} \over 2} p + {1 \over 2} ({1 \over 2} +
{1 \over \sqrt{2}})p^2+o(p^2).
\end{align}

Now, by continuity of $k$, $z(k)$ has the smallest absolute value in
a neighborhood of $k=0$.

Since $\partial_z D(k,z)|_{k=0,z=1} \neq 0$,  $z(k)$ has
multiplicity one. We then apply the Implicit Function Theorem to
find its derivatives.

\begin{rem}
For $p \rightarrow 1$, $D(0,z)\rightarrow 1-z,$ for all $z \in \{z:
|z|<{1 \over q}\}$, which implies that other roots go to infinity.
In the limit, when $p=1$, there is only a single root.
\end{rem}

Now we utilize the Implicit Function Theorem to find $\partial_k
z(0)$ and ${\partial_k}^2 z(0)$ where $z(k)$ is defined implicitly
by $D(k,z) \equiv 0$.

Taking the first derivative we get
\begin{equation}
\begin{split}
0=&\partial_k D(k,z) \\
 =&- p q \sin{k} z ^3 + p q \cos{k} 3 z^2 \partial_k z - (p q + p)
 2 z \partial_k z-\\
  &- p \sin{k} z + p \cos{k} \partial_k z +\\
  & + E (2 z \partial_k z + 2 \sin{k} z - 2 \cos{k} \partial_k z)
 +\\
  &+\partial_k E (z^2 - 2\cos{k} z +1).
\end{split}
\end{equation}

For $k = 0$ and $z = 1$, the equation becomes
\begin{equation}
(p q -p)\partial_k z = 0,
\end{equation}
which implies that
\begin{align}
\partial_k z (0)=0.
\end{align}
 Now, for ${\partial_k}^2 z(0)$, we can take the second derivative on $D(k,z)$
 to get
\begin{equation}
\begin{split}
0=& {\partial_k}^2 D(k,z)\\
 =& - p q \cos{k} z^3 -2 p q \sin{k} 3 z^2 \partial_k z+ \\
  & +p q \cos{k} (3 z^2 {\partial_k}^2 z+ 6 z (\partial_k z)^2)-\\
  & - (p q +p) (2 z {\partial_k}^2 z+ 2 (\partial_k z)^2) - p \cos{k} z -\\
  &- 2 p \sin{k} \partial_k z+ p \cos{k} {\partial_k}^2 z \\
  &+E (2 z {\partial_k}^2 z+2 \cos{k}z-2 \cos{k}
 {\partial_k}^2
 )\\
  & + 2 \partial_k E (2 z \partial_k z + 2 \sin{k} \partial_k z - 2 \cos{k} \partial_k
 z)\\
  & {\partial_k}^2 E(z^2-2 \cos{k}z+1),
\end{split}
\end{equation}
which in turn gives
\begin{align}
{\partial_k}^2 z (0)= {p + 2 \sqrt{1+q^2}-2 \over p}.
\end{align}
Similarly, taking the third derivative of $D(k,z)=0$ gives
\begin{align}
{\partial_k}^3 z (0)=0.
\end{align}
Also, by taking the fourth derivative we get
\begin{align}
\begin{split}
 {\partial_k}^4 z (0)=&{1 \over p^3 (1+q^2)^{1 \over
2}}(76q^4-83q^3(1+q^2)^{1 \over 2}+\\
&+16q^3+68q^2-q^2(1+q^2)^{1 \over 2}-37q(1+q^2)^{1 \over 2}+\\
&+16q-23(1+q^2)^{1 \over 2}+28).
\end{split}
\end{align}
 Hence we
have the expansion of $z(k)$ at $k=0$
\begin{align} \label{zkt}
z(k)= 1 + {p + 2 \sqrt{1+q^2}-2 \over 2 p} k^2 + O(k^4).
\end{align}

The residue of ${\widehat{P}(k,z) \over z^{t+1}}$ is
\begin{align} \label{res2}
Res({\widehat{P}(k,z) \over z^{t+1}}, z(k)) = ({1 \over z(k)})^{t+1}
\lim_{z \rightarrow z(k)}(z - z(k))\widehat{P}(k,z).
\end{align}

We now prove another lemma.
\begin{lem}
\begin{align}\label{res}
\lim_{z \rightarrow z(k)}(z(k) - z) \widehat{P} (k,z)=1+O(k^2)
\end{align}
as $k \rightarrow 0$.
\end{lem}
\emph{Proof}. Note that $\forall z \neq 1$,
\begin{align}
\lim_{k \rightarrow 0}(z(k) - z) \widehat{P} (k,z)=1,
\end{align}
i.e., $\forall \epsilon>0$, $\exists \delta$, s.t.,
\begin{align}\label{fbd}
|(z(k) - z) \widehat{P} (k,z)-1|<\epsilon
\end{align}
for $|k|<\delta$. (\ref{fbd}) implies that
\begin{align}
\lim_{z \rightarrow z(k)} |(z(k) - z) \widehat{P} (k,z)-1| \leq
\epsilon
\end{align} for $|k|< \delta$.
Hence,
\begin{align}
|\lim_{z \rightarrow z(k)} (z(k) - z) \widehat{P} (k,z)-1| \leq
\epsilon
\end{align}
for $|k|< \delta$, i.e.,
\begin{align}
\lim_{k \rightarrow 0} \lim_{z \rightarrow z(k)} (z(k) - z)
\widehat{P} (k,z)=1.
\end{align}
Now, for a small $r_1>0$ s.t. $z(k)$ is the only pole inside the
circle $|z-1|=r_1$, we have
\begin{equation}
\begin{split}
& \lim_{k \rightarrow 0} \lim_{z \rightarrow z(k)} {1 \over k}((z(k)
- z) \widehat{P} (k,z)-1)\\
=& \lim_{k \rightarrow 0}  {1 \over k}({1 \over 2 \pi
i}\oint_{|z-1|=r_1} \widehat{P}(k,z) d z - 1)\\
=& {1 \over 2 \pi i}\oint_{|z-1|=r_1}
\partial_k \widehat{P}(0, z) d z\\
=& 0.
\end{split}
\end{equation}
Similarly, we also have
\begin{equation}
\begin{split}
& \lim_{k \rightarrow 0} \lim_{z \rightarrow z(k)} {1 \over
k^2}((z(k)
- z) \widehat{P} (k,z)-1)\\
=& {1 \over 2 \pi i}\oint_{|z-1|=r_1}
{\partial_k}^2 \widehat{P}(0, z) d z\\
=& Res({z \over (1-z)^2} + {2 z^2 (-1+\sqrt{1+q^2z^2}) \over
(1-z)^2(p z + (1-z) \sqrt{1+q^2 z^2})}, 1)\\
=& {p+2\sqrt{1+q^2}-2 \over p} - {2 q^2 \over p \sqrt{1+q^2}} -{2
\over p^2} (1+q^2-\sqrt{1+q^2}).
\end{split}
\end{equation}
Therefore, $\lim_{z \rightarrow z(k)}(z(k) - z) \widehat{P}
(k,z)=1+O(k^2)$.

Now for any fixed $k \in [0, 2 \pi]$, the characteristic function of
${X_t \over \sqrt{t}}$ is $\widehat{P}({k \over \sqrt{t}},t)$. Since
the roots of $D(k,z)$ are isolated, we can set $r(p)=1+{\sqrt{2}
\over 2} p$ s.t. $|z({k \over \sqrt{t}})| < r(p)$ and other roots
are outside the circle $\{|z|=r(p)\}$. Furthermore, when $t$ is
large, ${k \over \sqrt{t}}$ is small, hence the lemmas are
applicable. We define the contour $C$ as $C = \{z: |z| = r_0\}
\bigcup \{z: |z|=r(p)\}$, where $r_0<1$.

By definition,
\begin{align}
 \widehat{P}({k \over \sqrt{t}},t)={1 \over 2 \pi i}\oint_{|z|=r_0} {\widehat{P}({k \over \sqrt{t}}, z) \over z^{t+1} } d z .
\end{align}
Since $z({k \over \sqrt{t}})$ is the only pole inside the contour,
we have
\begin{equation}
\begin{split}
 & -Res({\widehat{P}({k \over \sqrt{t}},z) \over z^{t+1}}, z({k
\over
\sqrt{t}})) \\
= &{1 \over 2 \pi i} (\oint_{|z|=r_0}
{\widehat{P}({k \over \sqrt{t}},z) \over z^{t+1}} d z - \oint_{|z|=r(p)} {\widehat{P}({k \over \sqrt{t}},z) \over z^{t+1}} d z).\\
\end{split}
\end{equation}
 For fixed $0 < p \leq 1$, $\sup_{k, |z|=r(p)}
|\widehat{P}({k \over \sqrt{t}},z)|< \infty.$ Hence,
\begin{align}
\oint_{|z|=r(p)} {\widehat{P}({k \over \sqrt{t}},z) \over z^{t+1}} d
z = O({r(p)}^{-t}).
\end{align}
We have
\begin{align}
\widehat{P}({k \over \sqrt{t}},t) = -Res({\widehat{P}({k \over
\sqrt{t}},z) \over z^{t+1}}, z({k \over \sqrt{t}})) + O({r(p)}^{-
t}).
\end{align}

Note that by (\ref{res}), we have

\begin{align}
\lim_{t \rightarrow \infty} \lim_{z \rightarrow z({k \over
\sqrt{t}})} (z({k \over \sqrt{t}}) - z) \widehat{P} ({k \over
\sqrt{t}},z) = 1 + O(t^{-1}).
\end{align}

Note also that by (\ref{zkt}),
\begin{align}
z({k \over \sqrt{t}})=1 + {p + 2 \sqrt{1+q^2}-2 \over 2 p} {k^2
\over t} + O( t^{-2}),
\end{align}
which implies that
\begin{equation}
\begin{split}
& ({1 \over z({k \over \sqrt{t}})})^{t+1}\\
=&( 1 + {p + 2 \sqrt{1+q^2}-2 \over 2 p} {k^2 \over t} + O(
t^{-2}))^{-(t+1)}\\
=&(1 - {p + 2 \sqrt{1+q^2}-2 \over 2 p} {k^2 \over t} + O(
t^{-2}))^{t+1}\\
=& \exp\{-{p + 2 \sqrt{1+q^2}-2 \over 2 p} k^2\} + O (t^{-1}),
\forall k.
\end{split}
\end{equation}
Therefore, by (\ref{res2}), $\forall k \in [0, 2 \pi]$,
\begin{align}
\widehat{P}(z({k \over \sqrt{t}}), t) = \exp\{-{p + 2 \sqrt{1+q^2}-2
\over 2 p} k^2\} + O (t^{-1}),
\end{align}
as $t \rightarrow \infty$. Hence, the limiting distribution of the
symmetric decoherent Hadamard walk is Gaussian with variance $v = {p
+ 2 \sqrt{1+q^2}-2 \over p}$.
\\
\\
\noindent \emph{Proof of Theorem \ref{thm2.5}.} For $X_t$, we can
also find its long-term variance. Let $C$ be the same contour as
before, when $t$ is large, $z=1$ is the closest root to 0 among all
that of the denominator of $\widehat{P}(k,z)$.

Note that
\begin{equation}
\begin{split}
 & -{\partial_k ^2} \widehat{P}(0,z)\\
=& {z \over (1-z)^2} + {2 z^2 (-1+\sqrt{1+q^2z^2}) \over (1-z)^2(p z
+ (1-z) \sqrt{1+q^2 z^2})},
\end{split}
\end{equation}
and that
\begin{equation}
\begin{split}
 &Res({1 \over z^{t+1}}({z \over (1-z)^2} + {2 z^2
(-1+\sqrt{1+q^2z^2}) \over
(1-z)^2(p z + (1-z) \sqrt{1+q^2 z^2})}), 1)\\
=& (-1-{2(\sqrt{1+q^2}-1) \over p}) t + {2 q^2 \over p \sqrt{1+q^2}}
+ {2 \over p^2} (1+q^2-\sqrt{1+q^2}).
\end{split}
\end{equation}
Hence,
\begin{equation}
\begin{split}
\!\!\!\!\! &Var(X_t)\\
=&-{\partial_k}^2 \widehat{P}(0,t)\\
=&-{1 \over 2 \pi i} \oint_C {{\partial_k}^2 \widehat{P}(0,z) \over
z^{t+1}} d
z\\
=&-Res({1 \over z^{t+1}}({z \over (1-z)^2} +\\
 &+ {2 z^2 (-1+\sqrt{1+q^2z^2}) \over
(1-z)^2(p z + (1-z) \sqrt{1+q^2 z^2})}), 1)+O(r(p)^{-t})\\
=&{p+2\sqrt{1+q^2}-2 \over p} t - {2 q^2 \over p \sqrt{1+q^2}}
-\\
 & -{2 \over p^2} (1+q^2-\sqrt{1+q^2})+O(r(p)^{-t}).
\end{split}
\end{equation}

The change of the order of integration and differentiation is
justified since ${\partial_k}^2 \widehat{P}(k,z)$ is continuous on
the contour. Hence, the long-term variance of the walk is
${p+2\sqrt{1+q^2}-2 \over p} t - {2 q^2 \over p \sqrt{1+q^2}} -{2
\over p^2} (1+q^2-\sqrt{1+q^2})+O(r(p)^{-t})$.

\section{Proofs in Subsection \ref{sit:2}}
From the decoherence equation we have
\begin{eqnarray}
\widehat{P}_{1,1}(k,z) &=& - {q \over p} + {1 \over p} {1-Q_{2,2}
\over
\det{(I-Q)}},\\
\widehat{P}_{1,2}(k,z) &=& {1 \over p} {Q_{1,2} \over \det{(I-Q)}}.
\end{eqnarray}
Let $\tilde{X_t}$ be the walk starting at position 0 with coin state
1 at time $t$ and $\widehat{\tilde{P}}(k,z) = \widehat{P}_{1,1}(k,z)
+ \widehat{P}_{1,2}(k,z)$ be its generating function. Then
\begin{equation}
\begin{split}
 &\widehat{\tilde{P}}(k,z)\\
=&-{q \over p} + {1 - \Sigma_1 + i \Sigma_3 + \Sigma_2 +
i \Sigma_4 \over p \det{(I-Q)}}\\
=&\widehat{P}(k,z) + i {\Sigma_3 + \Sigma_4 \over p \det{(I-Q)}}.
\end{split}
\end{equation}

Note that $\Sigma_3$ and $\Sigma_4$ both have a factor of $\sin{k}$,
we denote ${\Sigma_3 \over \sin{k}}$ and ${\Sigma_4 \over \sin{k}}$
by $\tilde{\Sigma_3}$ and $\tilde{\Sigma_4}$ respectively.\\
\\
\noindent \emph{Proof of Theorem \ref{thm2.6}.}
 Note that
\begin{equation}
\begin{split}
 & \partial_k \widehat{\tilde{P}} (0,z )\\
=& \partial_k \widehat{P}(0,z) + i \partial_k (\sin{k}
{\tilde{\Sigma_3} + \tilde{\Sigma_4}
\over p \det{(I-Q)}}) |_{k=0}\\
=& i ({{\tilde{\Sigma_3}} + \tilde{\Sigma_4} \over p}) |_{k=0}\\
=& i { z (\sqrt{1 + q^2 z^2} - 1) \over (1 - z) ( p z + (1 - z)
\sqrt{1 + q^2 z^2} )}.
\end{split}
\end{equation}

Let $C$ be the same contour as before. When $t$ is large, $z=1$ is
closest to 0 among the roots of the above denominator. Hence, for
fixed $p$,
\begin{equation}
\begin{split}
 & E(\tilde{X_t}) \\
=& {1 \over i} \partial_k \widehat{\tilde{P}}(0,t)\\
=& {1 \over 2 \pi i} \oint_C {\partial_k \widehat{\tilde{P}}(0,z)
\over i z^{ t
+ 1}} d z\\
=& {1 \over 2 \pi i} \oint_C {(\sqrt{1 + q^2 z^2} - 1) \over z^t (1
- z) ( p z + (1 - z) \sqrt{1 + q^2 z^2} )} d z\\
=& Res({(\sqrt{1 + q^2 z^2} - 1) \over z^t (1 - z) ( p z + (1 - z)
\sqrt{1 + q^2 z^2} )}, 1) + O({r(p)}^{- t}) \\
=& {\sqrt{1 + q^2 } - 1 \over p} + O({r(p)}^{- t}).
\end{split}
\end{equation}
\\
\noindent \emph{Proof of Theorem \ref{thm2.7}.} Note that
$-{\partial_k ^2} \widehat{\tilde{P}}(0,z)$ must be real. Hence,
\begin{equation}
-{\partial_k ^2} \widehat{\tilde{P}}(0,z)=-{\partial_k ^2}
\widehat{P}(0,z).
\end{equation}
Therefore, the formula is the same as before.\\
\\
\noindent \emph{Proof of Theorem \ref{thm2.8}.} Let $\mu_t =
E(\tilde{X_t})$. We want to show that ${\tilde{X_t} - \mu_t \over
\sqrt{t}} \rightarrow N(0,v)$. The long-term characteristic function
is
\begin{equation}
\begin{split}
 &\widehat{\tilde{P}}({k \over \sqrt{t}} ,t) e^{ - i {\mu_t k \over \sqrt{t}}}\\
=& \widehat{P}({k \over \sqrt{t}}, t) e^{ - i {\mu_t k \over
\sqrt{t}}} + \\
 & + e^{ - i {\mu_t k \over \sqrt{t}}} \sin{{k \over \sqrt{t}}} {1
\over 2
\pi i} \oint_C {1 \over z^{t+1}}({\tilde{\Sigma_3} + \tilde{\Sigma_4} \over p \det{(I-Q)}})({k \over \sqrt{t}}, z) d z\\
=&  \exp\{-{p + 2 \sqrt{1+q^2}-2 \over 2 p} k^2\} + O (t^{-{1 \over
2}}).
\end{split}
\end{equation}

Hence, the limiting distribution of the decoherent Hadamard walk
starting from 0 with coin state 1 is Gaussian with variance $v = {p
+ 2 \sqrt{1+q^2}-2 \over p}$.

\section{An Alternative Proof of Theorem \ref{pqv}:}
By Theorem \ref{fgf},
\begin{equation}\label{pvar}
\begin{split}
 & -{\partial_k ^2} \widehat{P}(0,z)\\
=& {z \over (1-z)^2} + {2 z^2 (-1+\sqrt{1+q^2z^2}) \over (1-z)^2(p z
+ (1-z) \sqrt{1+q^2 z^2})}. \end{split}\end{equation}

Note that the variance of the walk at time $t$ is given by
\begin{equation}
Var(X_t)=-\partial_k^2 \widehat{P}(0,t)
\end{equation}
and
\begin{equation}
\widehat{P}(0,z)= \sum_t \widehat{P}(0,t) z^t.
\end{equation}

Thus, $Var(X_t)$ is the $t$-th coefficient of the Taylor expansion
of $-\partial_k^2 \widehat{P}(0,z)$.

As $p \rightarrow 0$, $X_t$ becomes $Q_t$. In particular, the
function $V_0(z)=\sum_{t=1} Var(Q_t) z^t$ can be obtained  from
\ref{pvar},
\begin{align}
V_0(z)= {z \over (1-z)^2} + {2 z^2 \over (1-z)^3}(1-{1 \over
\sqrt{1+z^2}}).
\end{align}
Comparing the coefficients of the Taylor expansion of $V_0(z)$ gives
\begin{equation}
Var(Q_t)=t-\sum_{j=1}^{[{t-2 \over 2}]} (t-2j)(t-2j-1)(-1)^j ({1
\over 2})^{2j}{1 \over j} {(2j)! \over (j!)^2}.
\end{equation}
As $t \rightarrow \infty$,
\begin{equation}
\begin{split}
&  {Var(Q_t) \over t^2} \\
\rightarrow & - \sum_{j=1}^{\infty} (-1)^j ({1 \over 2})^{2j}{1
\over n} {(2j)! \over (j!)^2} \\
=& 1 - {1 \over \sqrt{2}}.
\end{split}
\end{equation}

\end{document}